\documentclass[prc,twocolumn,showpacs,amsmath,amssymb]{revtex4}
\usepackage{graphics,epsfig}
\makeatletter

\begin{document}
\title{Astrophysical Rate for $^{12}$N($p,\gamma$)$^{13}$O Direct Capture Reactions}
\author{LI Zhi-Hong}\email{zhli@ciae.ac.cn}
\affiliation{China Institute of Atomic Energy, PO Box 275(46),
Beijing 102413}

\begin{abstract}
The proton capture on the unstable nuclei plays a very important
role for the nucleonsynthesis. The $^{12}$N($p,\gamma$)$^{13}$O
reaction rates at the energies of astrophysical interests are
estimated with the spectroscopic factor and asymptotic normalization
coefficient methods. The present results show that the
$^{12}$N($p,\gamma$)$^{13}$O reaction may play an important role in
x-ray bursts.
\end{abstract}


\maketitle

Unstable nuclei play a very important role in the nucleonsynthesis.
If there are  not unstable nuclei but stable nuclei, most elements
will have not been synthesized due to the lack of reaction routes.
The proton- and $\alpha$-capture reactions on light nuclei close to
the proton drip line are very important in the evolution of super
massive stars. The studies on the energy release and the explosion
mechanism of the high-mass, low-metallicity stars (Population
I$\!$I$\!$I) are mainly dependent on their reaction rates.
Therefore, the determination of their reaction rates is of great
importance, particularly at higher burning temperatures.

The $^{12}$N unstable nucleus, which is produced via the
$^9$C($\alpha,p$)$^{12}$N and $^{11}$C($p,\gamma$)$^{12}$N
reactions, may capture a proton and form $^{13}$O via the
$^{12}$N($p,\gamma$)$^{13}$O reaction, if the capture rate is
sufficiently large as compared with the $\beta^+$-decay. The
$^{12}$N($p,\gamma$)$^{13}$O reaction is believed to play a role in
the nucleonsynthesis of Population I$\!$I$\!$I stars.\cite{Wie89} Up
to date, although the $^{12}$N($p,\gamma$)$^{13}$O reaction has been
discussed by many researchers,\cite{Wie89, Kaj90, Ter03, Min03} the
uncertainty of the $^{12}$N($p,\gamma$)$^{13}$O reaction rates is
still exist. Thus, it is necessary to determine the
$^{12}$N($p,\gamma$)$^{13}$O reaction rates at the temperatures of
astrophysical interest with an independent method.

\begin{figure}
\includegraphics[height=5.8 cm]{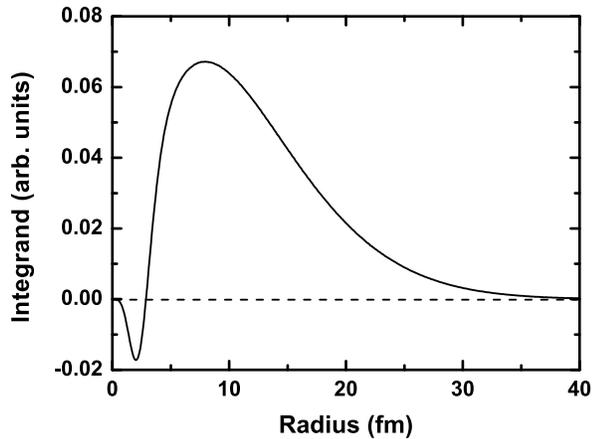}
\caption{\label{fig:integral} Integrand of the E1 transition matrix
element base on a single-particle model at 0.5\,MeV.}
\end{figure}

The proton separation energy of $^{13}$O is very low (1.51\,MeV).
The direct capture of $^{12}$N($p,\gamma$)$^{13}$O reaction is
believed to be dominated by the $E1$ transition from incoming
$s$-wave to bound $p$ state, the contribution of $d$-wave is
negligible at the energies of astrophysical interests. According to
the traditional direct capture model,\cite{Rol73} the cross section
of proton direct capture to the ground state of $^{13}$O with the
orbit and total angular momenta $l_f$ and $j_f$ can be expressed as
\begin{eqnarray}\label{eq1}
\sigma_{_{DC}} = {16 \pi \over 9}k_{\gamma}^{3}{e_{eff}^{2} \over
k^{2} }{1 \over \hbar v}{(2I_{3}+1) \over
     (2I_{1}+1)(2I_{2}+1) }S_{l_{f},j_{f}}\nonumber\\
 \times|\int^{\infty}_{0}r^2 \psi_{l_i}(kr)\phi_{l_f}(\kappa_B r)dr|^{2} ,
\end{eqnarray}
where $k_{\gamma}=\epsilon_{\gamma}/\hbar c$ is the wave number of
the emitted $\gamma$-ray (of energy $\epsilon_{\gamma}$); $e_{eff} =
eN/(A+1)$ is the proton effective charge for the E1 transition in
the potential produced by a target nucleus with mass number $A$ and
neutron number $N$; $v$ corresponds to the relative velocity between
$^{12}$N and proton; $k$ = $\sqrt{2\mu E_{cm}}/\hbar$ is the wave
number of the incident proton; $I_1$, $I_2$, and $I_3$ are the spins
of proton, $^{12}$N, and $^{13}$O, respectively; $S_{l_f,j_f}$ is
the spectroscopic factor of the configuration $^{13}$O $\rightarrow$
$^{12}$N + $p$; $\psi_{l_i}(r)$ is the optical model scattering wave
function of the colliding proton and $^{12}$N; and
$\varphi_{l_f}(r)$ is the radial wave function of the bound state
$p+^{12}$N in $^{13}$O, which can be calculated by solving the
respective Schr\"odinger equation with the optical potential model.
If the spectroscopic factor $S_{l_f,j_f}$ is known, the
$^{12}$N($p,\gamma$)$^{13}$O cross section can then be calculated by
Eq.\,(1).

\begin{figure}
\includegraphics[height=5.8 cm]{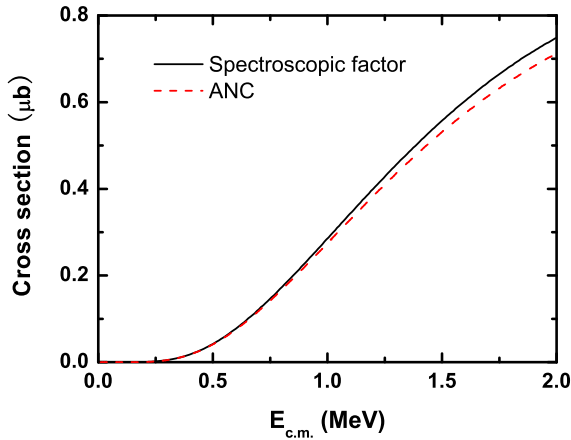}
\caption{\label{fig:section} Cross section of the
$^{12}$N($p,\gamma$)$^{13}$O direct capture reaction calculated with
Eqs.\,(1) and (2), respectively.}
\end{figure}

The integrand of the E1 transition matrix element based on a
single-particle model at 0.5\,MeV is shown in Fig.\,1, One can see
that the contribution to the $^{12}$N($p,\gamma$)$^{13}$O direct
capture reaction at small $r$ is not significant. In this case, the
$^{12}$N($p,\gamma$)$^{13}$O direct capture reaction is dominated by
the peripheral process. The cross section of peripheral proton
capture is not sensitive to the optical potential parameters and can
be calculated with the asymptotic normalization coefficient (ANC)
method and Eq.\,(1) can be replaced by
\begin{eqnarray}\label{eq2}
\sigma_{_{DC}} = {16 \pi \over 9}k_{\gamma}^{3}{e_{eff}^{2} \over
k^{2} }{1 \over \hbar v}{(2I_{3}+1) \over
     (2I_{1}+1)(2I_{2}+1) }C^2_{l_{f},j_{f}}\nonumber\\
 \times|\int^{\infty}_{R_N}r \psi_{l_i}(kr)W(2 \kappa_B r)dr|^{2} ,
\end{eqnarray}
where $C^2_{l_{f},j_{f}}$ is the squared proton ANC for $^{13}$O
ground state; $W(2\kappa_Br)$ is the Whittaker hypergeometric
function; $k_B$ is the bound state wave number for the last proton
in $^{13}$O; and $R_N$ is the interaction radius between proton and
$^{12}$N, which can be calculated by
\begin{equation}
\label{eq3}%
R_N = 1.25(A^{1/3}+1.0).
\end{equation}

The spectroscopic factor describes the overlap between the initial
and final states in the reaction channels and yields important
information about single-particle orbitals in many nuclei. It is
also an important ingredient for calculation of direct transfer
reaction cross sections in the distorted wave Born approximation
(DWBA) and capture reaction cross sections in the direct capture
(DC) model. Thus, great efforts have been expended in the studies of
spectroscopic factors theoretically and experimentally.
\cite{Cohen67,Ver94, Brown02,Tsang05} The valence proton can occupy
the $1p_{1/2}$ or $1p_{3/2}$ states in $^{13}$O. The spectroscopic
factors $S_{1,3/2}$ = 0.086 and $S_{1,1/2}$ = 0.537 are obtained,
based on the calculations of the shell model code
OXBASH.\cite{Brown84} These values are in good agreement with the
resent experimental data of the one proton removal cross
section.\cite{War06} The square of nuclear ANC is deduced to be
3.02\,fm$^{-1}$ based on the optical model with the standard
geometry parameters. Figure 2 shows the cross section of the
$^{12}$N($p,\gamma$)$^{13}$O direct capture reaction calculated with
Eq.~(\ref{eq1}) and (\ref{eq2}), respectively. The two results agree
with each other within 2\%, which demonstrates the identical result
from the two methods in the calculations of the
$^{12}$N($p,\gamma$)$^{13}$O direct capture reaction.

\begin{figure}
\includegraphics[height=5.8 cm]{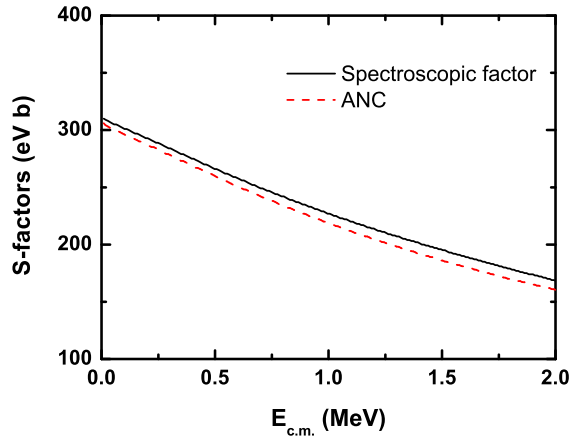}
\caption{\label{fig:sfactor} Astrophysical $S$-factors for the
$^{12}$N($p,\gamma$)$^{13}$O direct capture reaction.}
\end{figure}

For the astrophysically important class of charged-particle-induced
fusion reactions, the cross section of the fusion reaction drops
nearly exponentially with decreasing energy due to the tunneling
effect through the Coulomb barrier. The astrophysical $S$-factor is
often used to extrapolate the reaction data to lower energies in the
Gamow window. The $S$-factor is defined as
\begin{equation}
\label{eq4}%
S(E)=E\sigma(E)\exp(E_{G}/E)^{1/2},
\end{equation}
where the Gamow energy $E_{G}=0.978Z^{2}_{1}Z^{2}_{2}\mu$\,MeV,
$Z_{1,2}$ denote the atomic number of $^{12}$N and proton, $\mu$ is
the reduced mass of the system. Figure 3 shows the $S$-factors for
the $^{12}$N($p,\gamma$)$^{13}$O direct capture reaction as
functions of energies with the above-mentioned two methods. The
present $S$-factor at zero energy is 0.31\,keV\,b, which is about
two orders of magnitude less than the only theoretical one (4.0
$\times 10^{-2}$\,MeV\,b) in Ref.\,[1]. Further experimental works
are necessary to determine the cross section and $S$-factor of
$^{12}$N($p,\gamma$)$^{13}$O.

\begin{figure}[b]
\includegraphics[height=5.8 cm]{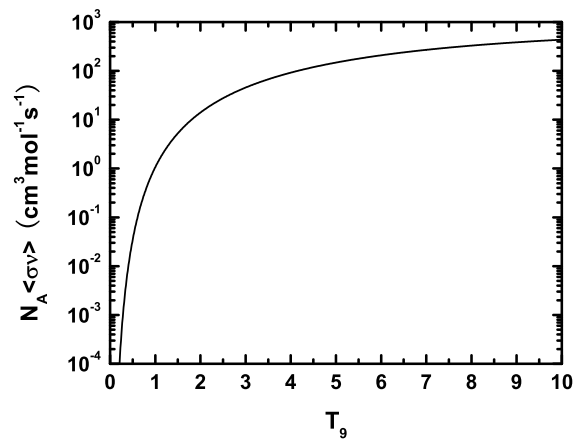}
\caption{\label{fig:rates} Rates of the $^{12}$N($p,
\gamma$)$^{13}$O direct capture reaction as a function of
temperature.}
\end{figure}

The astrophysical reaction rate of the $^{12}$N($p,\gamma$)$^{13}$O
direct capture can be calculated by
\begin{eqnarray}
\label{eq5}%
N_A \langle\sigma v\rangle & =&  N_A\big({8 \over
\pi\mu}\big)^{1/2}{1
\over (kT)^{3/2}}\nonumber\\
&\times&\int^{\infty}_0 S(E)\exp\big[-({E_{G} \over E})^{1/2}-{E
\over kT}\big]dE,
\end{eqnarray}
where $N_{A}$ is Avogadro's constant. The average value of the two
$S$-factors in Fig.\,3 is used in the present calculations.

Figure 4 shows the reaction rates as a function of temperature $T_9$
(in units of GK), the rates are fitted with an expression used in
the astrophysical reaction rate library REACLIB, \cite{Thi87}
\begin{eqnarray}
\label{eq5}%
N_A \langle\sigma v\rangle &=& \exp[15.1767+0.0037T_{9}^{-1}
-15.5324T_{9}^{-1/3}\nonumber\\& &+0.4554T_{9}^{1/3}-0.0318T_{9}
-0.0115T_{9}^{5/3} \nonumber\\& &-0.9821\ln(T_{9})].
\end{eqnarray}
The overall fitting errors are less than 2\% in the range from
$T_{9}$ =
0.01 to $T_{9}$ = 10.  The present reaction rates of $^{12}$N($p,\gamma$)$^{13}$O can be used in the nuclear reaction network calculations.\\

\begin{figure}
\includegraphics[height=6.0 cm]{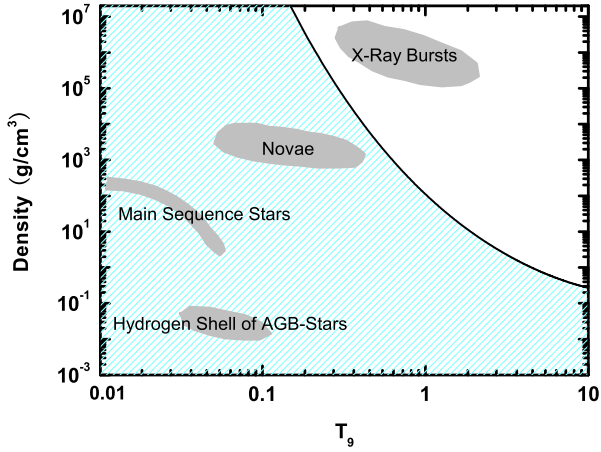}
\caption{\label{fig:density} Temperature and density boundary at
which the $^{12}$N($p,\gamma$)$^{13}$O direct capture reaction and
the competing $\beta^+$-decay are of equal strength assuming the
hydrogen mass fraction of $X_H$ = 0.77. The typical temperature and
density conditions for x-ray bursts, novae, Main Sequence stars and
the Hydrogen shell of AGB-stars are taken from Ref.\,[12].}
\end{figure}

\vspace{-2mm}The competition between the $^{12}$N($p,
\gamma$)$^{13}$O direct capture reaction and $^{12}$N
$\beta^+$-decay depends on the density, temperature and mass
fraction of proton in stars. The solid line in Fig.\,5 shows the
temperature and density boundary at which the
$^{12}$N($p,\gamma$)$^{13}$O direct capture reaction and the
competing $\beta^+$-decay are of equal strength assuming the
hydrogen mass fraction of $X_H$ = 0.77. The typical temperature and
density conditions for x-ray bursts, novae, Main Sequence stars and
the Hydrogen shell of AGB-stars taken from Ref.~\cite{Wie99} are
also shown in this figure. In the region above the solid curve the
proton capture reaction dominates, while below the solid line the
$^{12}$N nuclei are exhausted by the $\beta^+$-decay. Compared with
the typical temperature and density conditions for x-ray bursts,
novae, main sequence stars and the hydrogen shell of AGB-stars, one
can find that the $^{12}$N($p,\gamma$)$^{13}$O direct capture
reaction may play an important role in x-ray bursts.

\vspace{0mm} In summary, using the spectroscopic factor and ANC of
$^{13}$O ground state, the astrophysical rates of the $^{12}$N($p,
\gamma$)$^{13}$O direct capture reaction have been estimated at the
energies of astrophysical interests. The present results show that
the $^{12}$N($p,\gamma$)$^{13}$O reaction may play an important role
in x-ray bursts.

{\small This work is supported by the National Natural Science
Foundation of China under Grant Nos 10375096 and 10675173.}


\begin{thebibliography}{50}
\bibitem{Wie89} M. Wiescher {\it et al.}, 1989 {\em Astrophys. J.} {\bf 343} 352
\bibitem{Kaj90} T. Kajino and R.N. Boyd, 1990 {\em Astrophys. J.} {\bf 359} 267
\bibitem{Ter03} T. Teranishi {\it et al.}, 2003 {\em Nucl. Phys.} A {\bf 718} 207c
\bibitem{Min03} T. Minemura {\it et al.}, 2003 {\it Proceedings of the Third International Conference on Exotic Nuclei and Atomic Masses}
(Berlin: Springer) p 183
\bibitem{Rol73} C. Rolfs, 1973, {\em Nucl. Phys.} A {\bf 217} 29
\bibitem{Cohen67} S. Cohen and D. Kurath, 1967 {\em Nucl. Phys.} A {\bf 101} 1
\bibitem{Ver94} J. Vernotte {\it et al.}, 1994 {\em Nucl. Phys.} A {\bf 571}
1
\bibitem{Brown02} B. A. Brown, 2002, \prc {\bf 65}, 061601(R)
\bibitem{Tsang05} M. B. Tsang, J. Lee and W. G. Lynch, 2005, \prl {\bf 95}, 222501
\bibitem{Brown84} B. A. Brown, A. Etchegoyen and W.D.M. Rae, 1984 {\em code OXBASH} unpublished
\bibitem{War06} R. E. Warner {\it et al.}, \prc {\bf 74} 014605 (2006).
\bibitem{Wie99} M. Wiescher, J. G\"orres and H. Schatz {\it et al.}, {\em J. Phys.} G {\bf 25},
R133 (2006).
\bibitem{Thi87} F. K. Thielemann, M. Arnould and J. Truran 1987 {\it Advances in Nuclear Astrophysics} ed Vangioni-Flam E et al (Gif-sur-Yvette: Editions Frontieres)
\end{thebibliography}
\end{document}